\documentclass[preprint2,longabstract,eqsecnum]{aastex}

\usepackage{natbib}
\usepackage{bm}

\def\arcm{\ifmmode {^{\scriptscriptstyle\prime}}
          \else $^{\scriptscriptstyle\prime}$\fi}
\def\pdeg{\ifmmode $\setbox0=\hbox{$^{\circ}$}\rlap{\hskip.11\wd0 .}$^{\circ}
          \else \setbox0=\hbox{$^{\circ}$}\rlap{\hskip.11\wd0 .}$^{\circ}$\fi}

\shorttitle{A planet or just noise?}
\shortauthors{J. Golomb, G. Rocha, {\it et al.}}

\begin{document}

\title{{\tt PlanetEvidence}: Planet or Noise?}


\author{Jacob Golomb}
\affil{University of Maryland, College Park, MD, 20742, USA}

\author{Gra\c{c}a  Rocha}
\affil{Jet Propulsion Laboratory, California Institute of Technology, 4800 Oak Grove Dr, Pasadena, CA 91109, USA}
\affil{California Institute of Technology, MC 249-17, 1200 East California Blvd, Pasadena, CA 91125, USA}

\author{Tiffany Meshkat}
\affil{Infrared Processing and Analysis Center, MC 100-22 or MC 314-6, 1200 East California Blvd, Pasadena, CA 91125, USA}

\author{Michael Bottom}
\affil{Jet Propulsion Laboratory, California Institute of Technology, 4800 Oak Grove Dr, Pasadena, CA 91109, USA}

\author{Dimitri Mawet}
\affil{California Institute of Technology, MC 249-17, 1200 East California Blvd, Pasadena, CA 91125, USA}

\author{ Bertrand Mennesson}
\affil{Jet Propulsion Laboratory, California Institute of Technology, 4800 Oak Grove Dr, Pasadena, CA 91109, USA}

\author{Gautam Vasisht}
\affil{Jet Propulsion Laboratory, California Institute of Technology, 4800 Oak Grove Dr, Pasadena, CA 91109, USA}

\author{Jason Wang}
\affil{California Institute of Technology, MC 249-17, 1200 East California Blvd, Pasadena, CA 91125, USA}
\affil{51 Pegasi b Fellow}

\email{golombj@umd.edu}
\email{graca@caltech.edu; graca.m.rocha@jpl.nasa.gov}


\begin{abstract}

The work presented here attempts at answering the question: how do we decide when a given adetection is a planet or just residual noise in exoplanet direct imaging data?
To this end we present a method implemented within a Bayesian framework: (1) to unify 'source detection', and, 'source characterization' into one single rigorous mathematical framework; (2) to enable an adequate hypothesis testing given the S/N of the data; (3) to enhance the detectability of planets faint signal in the presence of instrumental and background noise and to optimize the characterization of the planet. 
As a proof of concept we implemented a routine named {\tt PlanetEvidence} that integrates the nested sampling technique (Multinest)  with a post-processing technique, the Karhunen-Loeve Image Processing (KLIP), algorithm. 
This is a first step to recast such post-processing method into a fully Bayesian perspective. 
We test our approach on real direct imaging data, specifically using GPI data of $\beta$ Pictoris b, and, on synthetic data. 
We find that for the former the method strongly favors the presence of a planet (as expected) and recovers the true parameter posterior distributions.
While for the latter case our approach allows us to detect (true) dim sources invisible to the naked eye as real planets, rather than background noise, and set a new lower threshold for detection at the $2 \sigma$ level approximately. Further it allows us to quantify our confidence that a given detection is a real planet and not just residual noise (for example residual speckles).
The next natural step is to extend this approach to construct a Bayesian-based algorithm for blind detection, that is, not requiring an initial guess as to the location of the planet.
This is the subject of ongoing work.\\

\end{abstract}


\section{Introduction}

Detection and characterization of extrasolar planets has become a flourishing field in the last two decades or so.
Several detection techniques have been applied: indirect such as 'Radial Velocity' (RV), 'Transit' detection;  Microlensing, 
and direct: e.g. 'Direct Imaging' techniques. The latter offers a unique way to study exoplanets in the context
of their formation and evolution. While indirect techniques have resulted in the discovery of thousands of
planets (for example transit detections have produced by far the most results (thanks to the Kepler mission)),
 direct imaging has discovered only a handful of planets. The difficulty arises from the residual glare of
starlight at small orbital separations where most planets are expected to reside, due to diffraction, scattered
light, and speckles caused by defects in the optical system. One measure of the detectability of a planet is the
"raw contrast" (hereafter referred to as 'contrast'), defined as the ratio of the average starlight irradiance in a
region of interest to the average irradiance of unblocked starlight in an equivalent aperture centered on the
star.  
The problem resides in the resolution and contrast needed to confirm the presence of a planet in the image (\citet{oppenheimer09}).

The Coronagraph Instrument (CGI) on the Wide-Field Infrared Survey Telescope (WFIRST) will aim to reach raw contrasts of about 1e-9 to 1e-8 using stateof-
the-art starlight suppression and wavefront control techniques  in order to detect such planets.
A further contrast improvement of at least a factor of two is expected at the post-precessing step.

In order to reach such deep sensitivity limits, new instruments and image processing techniques have been developed. 
For example coronagraphs are used to block light from the star, thus suppressing the noise from the Point Spread Function (PSF) of the star, while leaving the planet visible. 
Its optics allow us to reach smaller inner working angles, but are affected by the stellar speckles which can dominate the flux
from the plane. To correct for this \citet{bottom17} proposed a coronagraphic Phase-Shifting interferometry based on Coherent differential imaging that applies coherence properties of the light to detect substellar companions.

At the post-processing stage there are a number of Image processing techniques that aim at modeling and
subtracting the stellar PSF, to allow the planet to become detectable, in effect increasing
the contrast achievable next to a bright star. Techniques such as: Angular Differential Imaging, ADI, (\citet{marois08}); Locally Optimized Combination of Images, LOCI, (\citet{lafreniere07}); Reference Differential Imaging, RDI; Spectral Differential Imaging, SDI, (\citet{marois00});
Principal Component Analysis, PCA, (\cite{amara12}) models the temporal variation of
the PSF by identifying the main linear components of the temporal variation, or KLIP (\citet{soummer12})
which uses the Karhunen-Loeve (KL) transform to model the PSF. 
It has been shown that PCA-based methods achieve greater sensitivity and that varying and optimizing the number of
principal components is one of the best ways to enhance the planet signal \citep{meshkat14}. 
Other techniques have been developed as well though not fully exploited yet, such as stochastic speckle discrimination, SSD, (\cite{gladysz08}), enhanced faint companion photometry and astrometry
using wavelength diversity (\cite{burke10}).

While these methods enhance the detectability of faint
astrophysical signals, they do generally create systematic biases in their observed properties. To tackle this
issue, KLIP-FM (\cite{pueyo16}), perturbation-based KL Image Processing Forward Modeling has been
developed. More recently, \citet{ruffio17,ruffio18} developed a Forward Model Matched Filter (FMMF)
which uses the forward model provided by KLIP-FM as the template of the matched filter to boost the signal from the planet and hence enhance its detectability.

PSF subtractions on both ground-based and space-based instruments have not yet
achieved the contrast gain needed to detect planets with masses lower than 1 Jupiter mass at
separations smaller than 0.1". 
Further study is required to enhance the detectability of planets faint
signals hidden in the instrumental and background noise as well as to improve their characterization i.e. render
an unbiased estimation of their Position (astrometry) and Intensity (photometry) and accurately estimate the
planet's parameter uncertainties. Another aspect of major importance is to ensure that the results are not
dependent on the post-processing method employed. It is indeed expected that the diverse set of post-processing
techniques would  give rise to similar results. However this is not necessarily the case as the signal-to-noise and self-subtraction of a detected point source may vary significantly depending on the technique. 

We present here a method implemented within a Bayesian framework: (1) to unify 'source detection', i.e. deciding whether a certain signal is due to a source, and, 'source characterization', i.e. determining the parameters of the source, such as position, flux or intensity, 
into one single rigorous mathematical framework; (2) to enable an adequate hypothesis testing given the S/N of the data; (3) to enhance the detectability of planets faint signal in the presence of instrumental and background noise and to optimize the characterization of the planet (i.e. its flux and astrometry).
As a proof of concept we developed and implemented a module named {\tt PlanetEvidence} that integrates the nested sampling technique (Multinest) (\citet{skilling04, skilling04b}) with a post-processing technique, the Karhunen-Loeve Image Processing (KLIP) (\citet{soummer12}).
For the latter we use a a python library for direct imaging of exoplanets and disks, {\tt pyKLIP} (\citet{wang15}\footnote {for details visit: \\ \url{https://pyklip.readthedocs.io/en/latest/}}), 
which uses an implementation of KLIP and KLIP-FM to perform point PSF subtraction. This is a first step to recast such post-processing method into a fully Bayesian perspective.

Our future  implementation of a blind detection step in both coadded and un-coadded data differ from the Forward Model Matched Filter (FMMF) module in {\tt pyKLIP} described in \citet{ruffio17,ruffio18}. 
We will not use matched filters to boost the signal from the planet but rather incorporate the noise (both white and correlated noise) in the covariance matrix of the Likelihood (assumed a Multivariate Gaussian) and marginalize over the nuisance parameters.
Our procedure follows the principles laid out in \citet{rocha09,rocha12}.

\citet{rocha09} have shown that maximizing the likelihood ratio between the two hypothesis, in the absence of the cross-term (see eq 16 in \citet{rocha12}), with respect to the source amplitudes $A_{j}$, recovers the expression for the Matched Filter, MF. This means that the filtered field is merely the projection of the likelihood manifold on to the sub-space of position parameters $X_{j}$. Hence they show that in the traditional approach to catalogue making, in which one compares the maximum SNR of the putative detections to some threshold, one is really performing a generalized likelihood ratio test. Furthermore, they lay out the foundations for a Bayesian-based generation of catalogs of point-sources. In this Bayesian method, the threshold is a byproduct of the method rather than set up a-priori (empirically), as is the case for frequentist-based methods.

In Section 2. we give an overview of KLIP-FM: Detection \& Forward Modeling while in Section 3. we give an overview of Nested sampling and our implementation in {\tt PlanetEvidence}.
In Section 4.  we present results \& discussion for both $\beta$ Pictoris b and synthetic data. Finally in Section 5. we present our conclusions.

\section{KLIP-FM: Detection \& Forward Modeling - an overview}

We extended an existing python library for direct imaging of exoplanets, {\tt pyKLIP} (\citep{wang15} \footnote {for details visit: \\ \url{https://pyklip.readthedocs.io/en/latest/}}), to the first stage of a fully Bayesian blind detection step. 
	
We used the KLIP algorithm to do stellar PSF subtraction (\citet{soummer12}). 
A companion algorithm to the KLIP method,  the KLIP Forward Modeling (KLIP-FM), allows for more accurate point-source detection using forward modeling (\citet{pueyo16}). KLIP-FM uses an input instrumental PSF at a given location and feeds it into the KLIP algorithm, thus resulting in a model instrumental PSF that can be corrected for biases due to PCA subtraction.  This results in a model that has accounted for self- and over-subtraction of a planet signal in the guessed position of the planet. The initial instrumental PSF model that is propagated through KLIP is unique to the instrument used. 

Here we make use of data from the Gemini Planet Imager (GPI) on the Gemini telescope (\citet{macintosh14}). For the GPI data used, the instrumental PSF is constructed using satellite spots (manually-projected point sources) of known locations in the image. 
We use the J-band coronagraph observations of $\beta$ Pictoris b, an exoplanet with a model dependent mass of $~10 M_{\rm Jup}$ at an orbital separation of $8 - 10$ AU (\citet{lagrange09, lagrange10}). 
These observational data, along with those for several other targets, are available at the Gemini data release website \footnote{ \url {https://www.gemini.edu/sciops/instruments/gpi/historical-documents/public-data}}. The $\beta$ Pictoris data come in the form of 19 time-series FITS datacubes, each containing 37 slices corresponding to exposures at different wavelengths ranging from $1.114 - 1.3497$ microns.

The {\tt pyKLIP} algorithm renders a KLIP-subtracted annulus that is plus and minus a user-prescribed number of pixels in the outward and inward radial directions from a guessed location of the planet. The guessed location of a planet is previously determined by eye from looking at the KLIP-subtracted image and looking for the presence of a planet. It also outputs another image, the post-KLIP instrumental PSF in the guessed location of the planet. This is then used as the model for the planet PSF in the image.

We explored one of two ways of extending {\tt pyKLIP} to Bayesian detection and characterization: (i) starting with the coadded frames (from the set of temporal and wavelength frames) after subtracting the stellar PSF (speckles); (ii) starting with a joint analysis of all frames in time and wavelength. We present here approach (i) while approach (ii) will be completed in the future.

Following \citet{soummer12}, we estimate the KL modes (up to some pre-set number) and subtract them from the image. Figure 1 shows a PSF-subtracted image of $\beta$ Pictoris (left hand side). With a KLIP-subtracted image and the location of a potential planet in mind, we initialize our forward model in that location (in this case 30.1 pixels from the center and 212 degrees counter clockwise from north). 
The fitting area, $\mathcal{F}$, is a 13 x 13 pixel box centered on the guessed location.
For the detection step, we follow \citet{wang16}: consider a set of parameters to minimize the Gaussian-distributed residuals between the data and instrumental PSF model in given locations, while accounting for residual spatially-correlated background noise (i.e. speckles) in the image. With F as the forward model of the instrumental PSF and D as the data, the aforementioned residuals over the fitting region  $\mathcal{F}$  are defined as:
  
\begin{equation}
    R \equiv (D - \alpha F(x_{p}, y_{p}))_{\mathcal{F}}
\end{equation}
 
The parameters introduced in this residual expression are spatial coordinates of the central location of the planet PSF $(x_{p}, y_{p})$  as well as a flux scale parameter  $\alpha$  to scale up or down the flux of the model to best match the data in  $\mathcal{F}$ .

The Likelihood function introduced by \citet{wang16} also considers the correlated nature of residual noise in the KLIP-subtracted image. Such noise is accounted for in the Matern covariance function with $\nu=3/2$ as motivated by \citet{czekala15}. The covariance, C, between the $i_{th}$ and $j_{th}$ pixel is calculated as:
  
\begin{equation}
    C_{ij} = \sigma_{i}\sigma_{j} (1 + \frac{\sqrt{3}r_{ij}}{\ell})\exp({\frac{-\sqrt{3}r_{ij}}{\ell}})
\end{equation}

Where $\ell$  is the correlation length scale, or the expected size of the residual correlated noise, and $r_{ij}$ is the distance between the $i_{th}$ and $j_{th}$ pixels in the image. It should be noted that  $\ell \simeq \frac{\lambda}{D}$ (~3 pixels in our test case), but it is still treated as a parameter and allowed to vary. Also, $\sigma_{i}$ is the uncertainty associated with the $i_{th}$ pixel, computed by taking the standard deviation of pixel values in an annulus containing the $i_{th}$ pixel. The likelihood function is a Gaussian of the residuals with the aforementioned covariance matrix, giving the log likelihood:
  
\begin{equation}
    ln\mathcal{L} = -\frac{1}{2}(R^{T}C^{-1}R + ln(det(C)) + N_{pix}ln(2\pi))
\end{equation}

\section{Nested Sampling: an overview and Implementation in {\tt PlanetEvidence}}

As opposed to getting the marginal distributions for each parameter using Markov-Chain Monte Carlo (MCMC) sampling as done in \citet{wang16}, we implement a nested sampling routine (introduced in \citet{skilling04, skilling04b}). Nested sampling was developed primarily to estimate the {\it evidence} (the average of the likelihood over the prior)
 for models being tested. As a byproduct it also provides the posterior distribution of the model parameters.
Furthermore this sampling has a great overall speed of computation.

With the equation for the probability of a set of $n$ parameters, $\theta_{n}$, given data, D, as simply:

\begin{equation}
    P(\theta_{n}|D) \propto \mathcal{L}(D|\theta_{n})\pi(\theta_{n})
\end{equation}

The expectation value of the likelihood with the priors $\pi (\theta_{n})$ is this equation integrated over all of the parameters. This is called the evidence term, and is calculated by:
  
\begin{equation}
    Z=P(D) = \int \mathcal{L}(D|\theta_{n})\pi(\theta_{n})d\theta_{n}
    \label{eq:Z}
\end{equation}

Given that for many parameter in a model, this equation can become difficult or impossible to calculate, nested sampling is used to perform the calculation. In nested sampling, this multidimensional integral over parameter space is transformed into a one-dimensional integral over probability space. Specifically, a function  $L(\lambda)$  is constructed, which is the probability of getting a likelihood of $\theta$ greater then $\lambda$ when sampling from the prior. The way this works is: N samples are made in prior space and the corresponding likelihood values are calculated. The lowest likelihood value is then {\it banked} (stored). It can therefore be said that there is an  $\frac{N-1}{N}$ probability of getting that likelihood or greater when sampling from prior space. This is done iteratively, each time storing the lowest value likelihood value. Each iteration only samples from prior space that corresponds to likelihoods greater than the previously-stored lowest likelihood value, building up a function $L(\lambda)$, the integral of which is calculated simply by summations. The equivalence between the integral over this probability space and that of the Likelihood times the prior (equation~\ref{eq:Z}) is explained in \citet{skilling04, skilling04b}.
 In addition, each parameter sample is assigned a corresponding {\it weight} to calculate marginal distributions for each parameter (for a detailed account see \citet{skilling04, skilling04b}). 
With this evidence term, we get an expectation value for the likelihood of the model given the data. 
It is worth noting that individual evidence values in its own right have no particular discriminatory meaning. It is only when the evidences for the two competing models $H_{0}$ and $H_{1}$ computed for the same data over the same fitting region are compared that one can draw conclusions.

For our purposes, we consider two models: one for the planet being present in the image, which we will call $H_{1}$, and another one for just the noise present in the image, $H_{0}$  (null hypothesis).
For $H_{1}$ model, we use the forward model with the same {\it Matern} covariance function to account for correlated noise. When we do sampling with this model, we are therefore looking for a point source, within a fitting region, that minimizes the residuals. For $H_{0}$  model, we assume there is no planet present in the fitting region, and we therefore set $\alpha=0$ and get the distributions for the remaining three parameters. Therefore, the Likelihood is simply the Likelihood of the data, with the $ \it Matern$ covariance to account for the correlated noise. The function for $H_{0}$ log Likelihood is therefore:  

\begin{equation}
    ln\mathcal{L} = -\frac{1}{2}(D^{T}C^{-1}D + ln(det(C)) + N_{pix}ln(2\pi))
\end{equation}

For the nested sampling implementation, we use {\tt pyMultiNest} (\citet{buchner14}), a python wrapper for the multimodal nested sampling algorithm, called MultiNest (\citet{feroz09}). This results in marginal distributions for all parameters, as well the evidence values $Z_{1}$ and $Z_{0}$ for the $H_{1}$ and $H_{0}$ hypothesis, respectively.
Noting that the expressions for $Z_{1}$  and $Z_{0}$ are:

\begin{equation}
    Z_{1} = \int_{\theta = \ell, x, y, \alpha} \mathcal{L}(D|\theta)\pi(\theta)d\theta
\end{equation}

and 

\begin{equation}
    Z_{0} = \int_{\theta = \ell,x,y} \mathcal{L}(D|\theta)\pi(\theta)d\theta
\end{equation}

We can perform Bayesian model comparison to determine how much one model is favored over the other. In particular, the evidence for $H_{1}$  over $H_{0}$ is simply given by the ratio of $Z_{1}$ to $Z_{0}$, ie $B_{10}=Z_{1}/Z_{0}$. Note that {\tt pyMultiNest} gives values for $lnZ$ rather than simply $Z$.
The confidence for which one model can be favored over another from the evidence ratio can be determined using, for example, the so-called "Jeffreys' scale" as tabulated in \citet{trotta08} or the so-called "Harold-Jeffreys" interpretation \citet{jeffreys61}.        

\section{Results \& Discussion}

\subsection{Results: Testing on $\beta$ Pictoris b}
  
We run our detection \& characterization routine, {\tt PlanetEvidence}, on real direct imaging data, specifically the GPI data.
We use the J-band coronagraph observations of $\beta$ Pictoris b (see Section 2. for more details).

\begin{figure}
  \begin{center}
  \hbox{
  \hspace*{0.1in}
    \includegraphics[width=0.4\columnwidth]{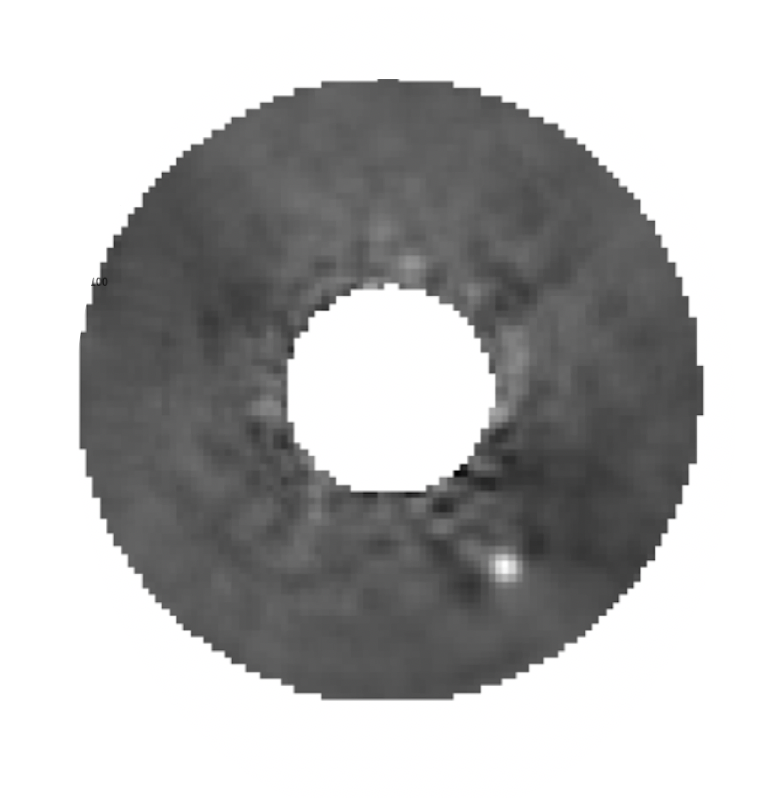}
      \includegraphics[width=0.4\columnwidth]{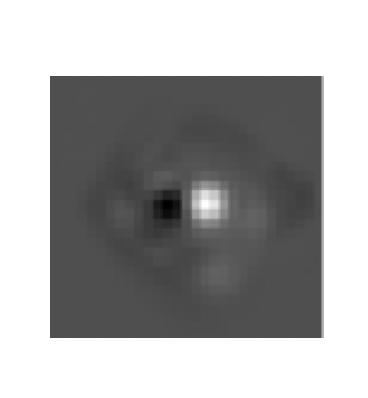}}
   \end{center}
  \caption{Left: KLIP-subtracted annulus from $\beta$ Pictoris b image. Note the presence of the planet in the 5 o'clock position. Right: An enlarged instrumental PSF forward modeled in the location of $\beta$ Pictoris b.}
  \label{fig:Pictorisb-I}
\end{figure}

We consider the guessed location of the planet in the image, at a radial separation of approximately 30.1 pixels from the center and at 212 degrees counterclockwise from the north of the image. 
We then use KLIP-FM to generate a forward model at that location. For the purposes of demonstration and consistency, we subtract 7 KL modes from the data, following the examples used for $\beta$ Pictoris b in {\tt pyKLIP-FM}. \footnote{\url{https://pyklip.readthedocs.io/en/latest/bka.html}}
See Figure~\ref{fig:Pictorisb-I} for a KLIP-subtracted annulus of $\beta$ Pictoris b data, as well as the forward model generated for that location. 
We use MultiNest with the forward model to get the posterior distributions for each of the four parameters in the $H_{1}$ model. 

The fitting area, $\mathcal{F}$, is a 13 x 13 pixel box centered on the guessed location. The priors for the central location of the planet's PSF is 1.5 pixels in both directions (x,y), while the priors for $\alpha$ and $\ell$ are uniform in log space with $\alpha$ between 0.016 and 1.58, and $\ell$ between 0.3 and 30 (as used in the example from {\tt pyKLIP} Forward modeling).

\begin{figure*}
  \begin{center}
   \includegraphics[width=2.0\columnwidth]{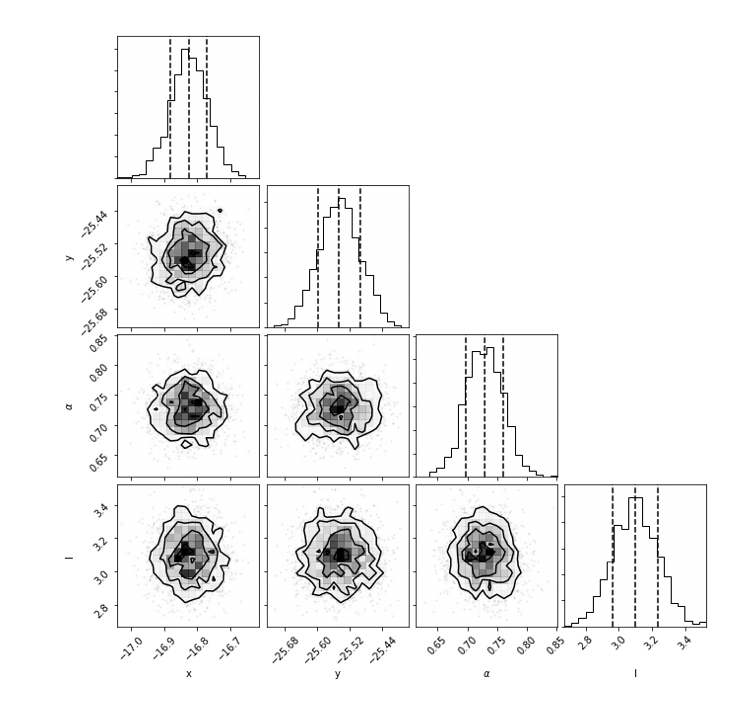}
   \end{center} 
   \vspace*{-0.4in}
   \caption{The posterior distributions of the four parameters of model $H_{1}$, $(x,y,\alpha,\ell)$.}
  \label{fig:Pictorisb-II-H1}
\end{figure*}

\begin{figure*}
  \begin{center}
    \includegraphics[width=2.0\columnwidth]{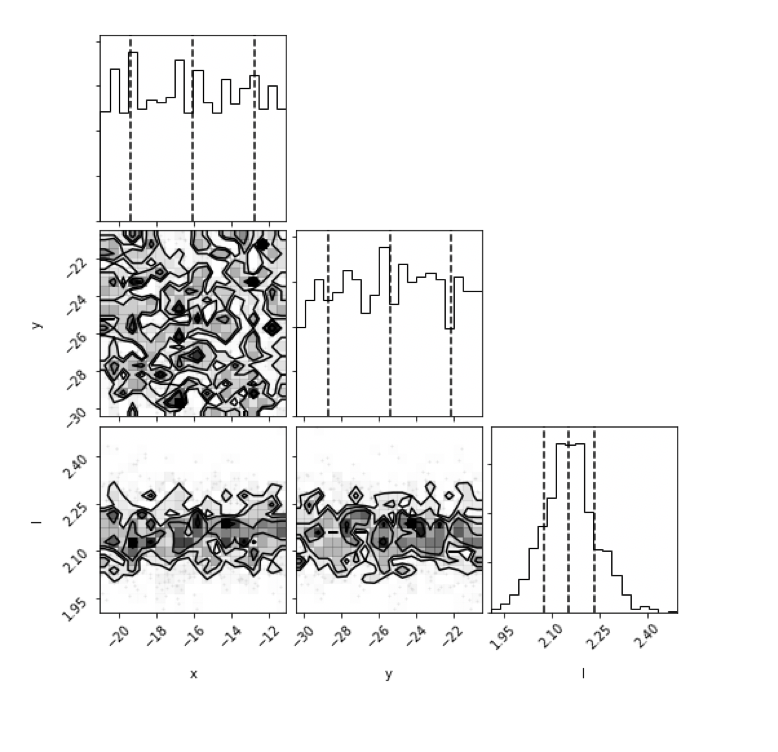}
   \end{center} 
   \vspace*{-0.6in}
   \caption{The posterior distributions of the three parameters of model $H_{0}$ (null hypothesis), $(x,y,\ell)$.}
  \label{fig:Pictorisb-II-H0}
\end{figure*}

\begin{figure*}
     \includegraphics[width=0.5\columnwidth]{beta_pictoris_b_annulus.png}
       \hspace*{0.1in}
    \includegraphics[width=1.55\columnwidth]{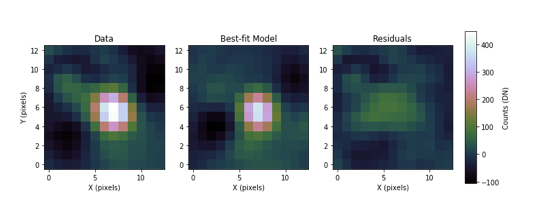}
   \caption{(Left) Annulus of KLIP-subtracted $\beta$ Pictoris b GPI data. (Right) three-panel plot: (left) search region around $\beta$ Pictoris b, (center) the best fit model for $H_{1}$ hypothesis and (right) the residuals after subtracting the best fit model.}
  \label{fig:Pictorisb-II-marg_bfm_res}
\end{figure*}

{\tt PlanetEvidence} renders the parameter distributions as well as the overall evidence for both $H_{0}$ and $H_{1}$ hypothesis. 
The marginal distributions of the model parameters, the best fit models, and the residuals are plotted in Figures~\ref{fig:Pictorisb-II-H1}, \ref{fig:Pictorisb-II-H0} and \ref{fig:Pictorisb-II-marg_bfm_res}.

Since $\beta$ Pictoris b is a bright planet at the guessed location in the image, we expect to recover parameter value, with low uncertainties, when fitting the forward model to the location of the planet in the image. This is clearly the case as shown in Figures~\ref{fig:Pictorisb-II-H1}. The distributions for the coordinates of the central position of the instrumental PSF, have subpixel uncertainties (on the order of +/- 0.05 pixels). Note that the positional parameters are measured as displacement from the center of the image, and manually checking the spot where these parameters converge to in the image reveals it converges to the center of the true location of $\beta$ Pictoris b. Also note that the $\alpha$ parameter converges to a value slightly less than 1, as our forward model PSF was slightly brighter than the true brightness of  $\beta$ Pictoris b. This is so because the forward model brightness is just an arbitrary guess of the planet flux. 

For $H_{0}$, the likelihood of just the data with the correlated residual noise is estimated along with the evidence for the null-hypothesis. The posterior distributions of the three parameters of model $H_{0}$ are shown in Figure~\ref{fig:Pictorisb-II-H0}.

The logarithm of the evidence ratios, $ln(B_{10}) = ln(Z_{1}/Z_{2}) \approx 175$, very strong evidence in favor of the $H_{1}$ model.
This is expected, as $\beta$ Pictoris b is a true, bright planetary signal in the image with a Signal to Noise ratio (SNR) of  $11.4 \sigma$. 

Note that we employ several methodologies to estimate the SNR of a source. First, we calculate $SNR_{bf,\mathcal{F}}$, where the signal is considered the peak pixel of the best-fit model, and the noise is the standard deviation of the residuals within the fitting area $\mathcal{F}$. Note that this calculation of SNR is thus sensitive to both the accuracy of the fit (if the best fit parameters are actually close to their true values) and the local residuals. For $SNR_{bf, an}$ we use the peak of the best-fit model as the signal and the standard deviation of the masked annulus as the noise. Finally,  $SNR_{an}$ is considered the brightest pixel at the injected location of the planet divided by the standard deviation of the planet-masked annulus ie the SNR of the planet as seen in the image after going through the KLIP step, while $SNR_{in,an}$ is the input flux of the planet with noise estimated in the whole annulus. Thus, for a PSF of a given flux injected into any angular position on the sky at the same separation, $SNR_{an}$ should be approximately the same, with differences due to, for example, the injected planet falling on a speckle.

For $\beta$ Pictoris b, we calculate $SNR_{an}=11.4\sigma$, and the SNR estimated from the residuals after subtracting the best fit model within the fitting zone, $SNR_{bf,\mathcal{F}}=11.5\sigma$. These SNRs, along with the Evidence ratio $ln(B_{10})=175$ are tabulated in table~\ref{beta Pictoris b}.
\begin{table*}
\caption{$\beta$ Pictoris b: SNR and Evidence ratios for models $H_{1}$ (planet model) and the null hypothesis, $H_{0}$. }
\begin{center}
\begin{tabular}{|c|c|c|c|c|}
\hline
Target & $SNR_{an}$ &$SNR_{bf,\mathcal{F}}$ &$|ln(B_{10})|$ & Strength of evidence \\
\hline
$\beta$ Pictoris b	&$11.4 \sigma $ 	& $11.5 \sigma $ 	&  175	& Very Strong\\
\hline
\end{tabular}
\end{center}
\label{beta Pictoris b}
\end{table*}

\subsection{Results: Testing on Synthetic Data}

To ascertain how dim a planet can be and still be detected using evidence ratios,  we create controlled cases by injecting synthetic planets into different locations in the image of $\beta$ Pic b, then running the detection and characterization routine. 

To inject a synthetic source we use {\tt pyklip.fakes} module of {\tt pyKLIP}. The fake sources are injected into the data before post-processing (i.e., before running the KLIP step) at the correct positions so that after the data are aligned and rotated, the fake planet will be aligned. 
\footnote {for details visit: \\ \url{https://pyklip.readthedocs.io/en/latest/contrast_curves.html/Injecting-fake-planets}}

\begin{figure*}
  \begin{center}
       \includegraphics[width=2.0\columnwidth]{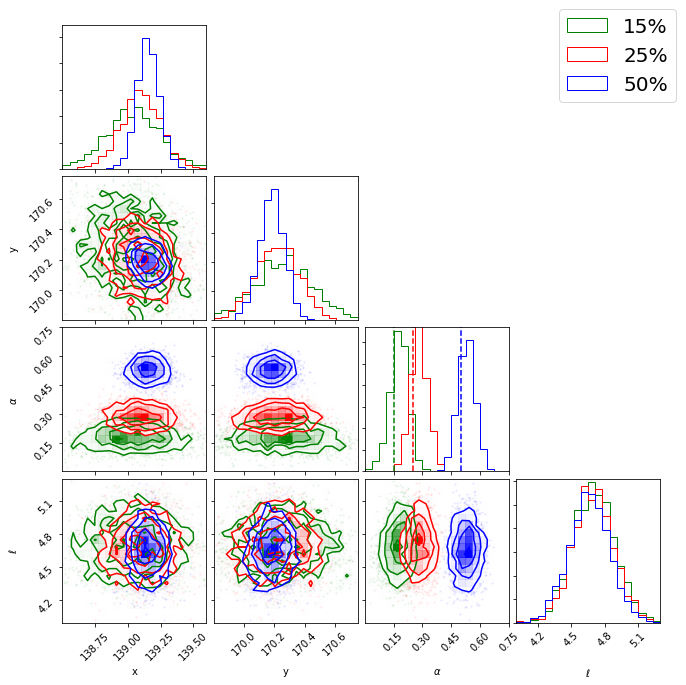}
   \end{center}
  \caption{The posterior distributions of the four parameters of model $H_{1}$, $(x,y,\alpha,\ell)$ for a planet injected at $pa = 0 \deg$ (12 o'clock) position with $50\%$ (blue), $25\%$ (red) and $15\%$ (green) of the Forward Model flux ($5 \times 10^{-5}$).}
  \label{fig:Pictorisb-mock_12_marg_H1}
\end{figure*}

\begin{figure*}
  \begin{center}
       \includegraphics[width=2.0\columnwidth]{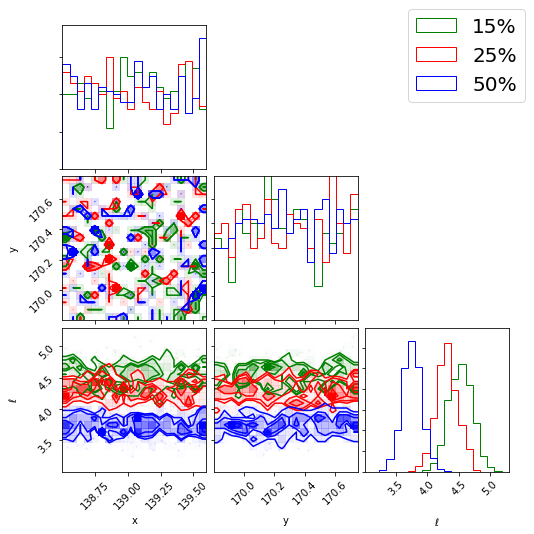}  
   \end{center}
  \caption{The posterior distributions of the three parameters of model $H_{0}$ (null hypothesis), $(x,y,\ell)$, for a planet injected at $pa = 0 \deg$ (12 o'clock) position with $50\%$ (blue), $25\%$ (red) and $15\%$ (green) of the Forward Model flux ($5 \times 10^{-5}$).}
  \label{fig:Pictorisb-mock_12_marg_H0}
\end{figure*}

\begin{figure*}
        \includegraphics[width=0.5\columnwidth]{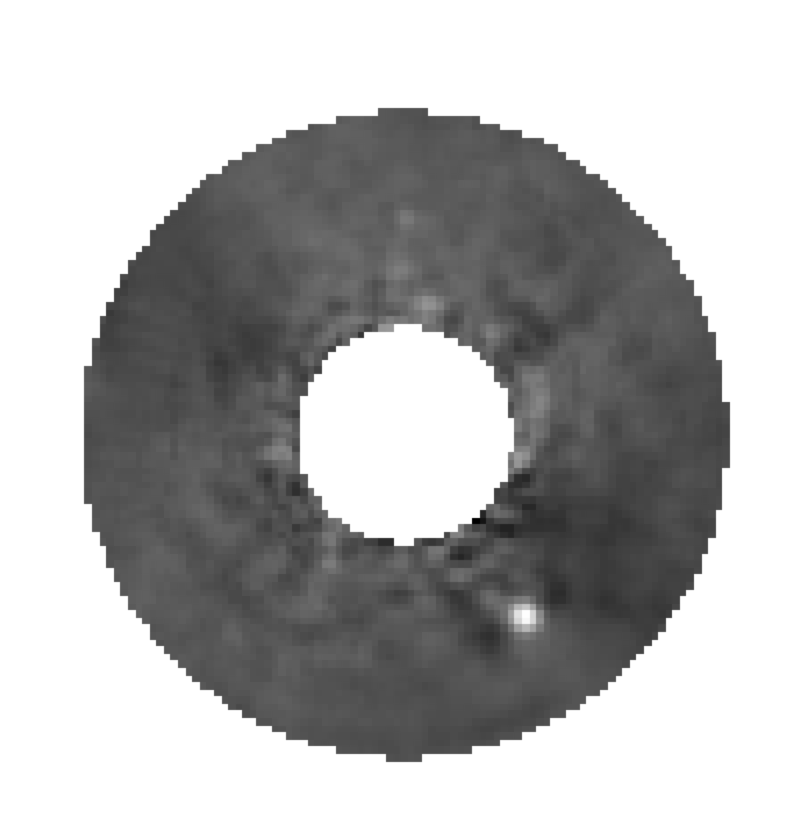}
        \hspace{0.2in}
        \includegraphics[width=1.5\columnwidth]{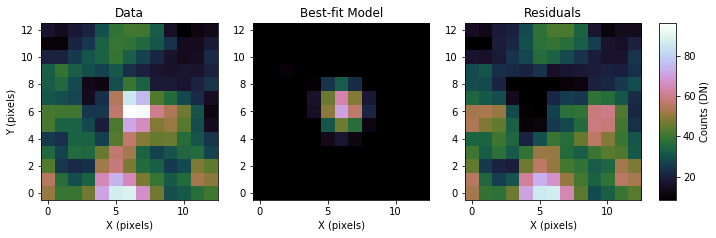}
  \caption{(Left) Annulus of KLIP-subtracted $\beta$ Pictoris b GPI data with a synthetic planet injected at $pa = 0 \deg$ position with $15\%$ of the Forward Model flux. (Right) three-panel plot: (left) search region around the injected planet, (center) the best fit model for $H_{1}$ hypothesis and (right) the residuals after subtracting the best fit model.}
  \label{fig:Pictorisb-mock_12_marg_bfm_res}
\end{figure*}

\begin{figure*}
  \begin{center}
       \includegraphics[width=2.0\columnwidth]{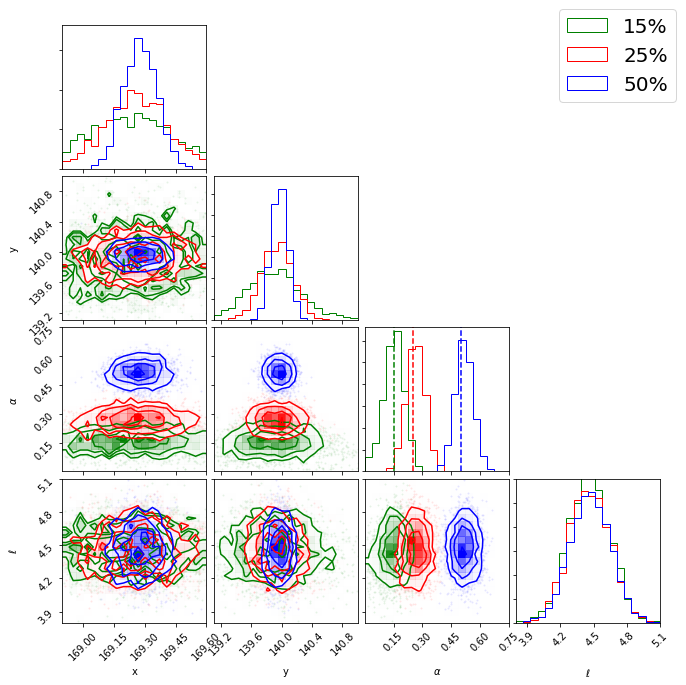}
   \end{center}
  \caption{The posterior distributions of the four parameters of model $H_{1}$, $(x,y,\alpha,\ell)$, for a planet injected at $pa = 270 \deg$  (3 o'clock) position with $50\%$ (blue), $25\%$ (red) and $15\%$ (green) of the Forward Model flux.}
  \label{fig:Pictorisb-mock_3_marg_H1}
\end{figure*}

\begin{figure*}
  \begin{center}
       \includegraphics[width=2.0\columnwidth]{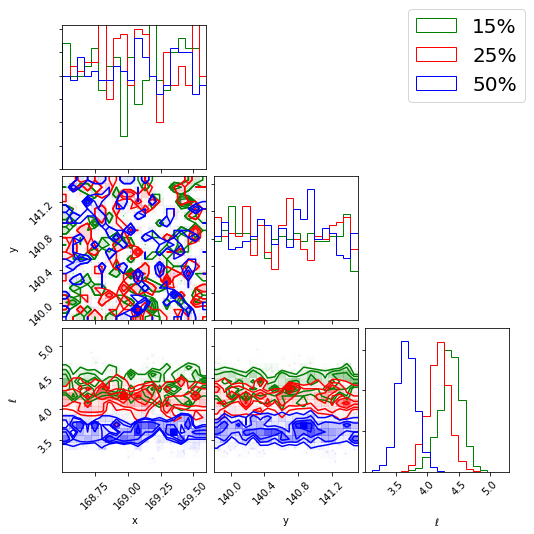}    
   \end{center}
  \caption{The posterior distributions of the three parameters of model $H_{0}$ (null hypothesis), $(x,y,\ell)$, for a planet injected at $pa = 270 \deg$  (3 o'clock) position with $50\%$ (blue), $25\%$ (red) and $15\%$ (green) of the Forward Model flux.}
  \label{fig:Pictorisb-mock_3_marg_H0}
\end{figure*}

\begin{figure*}
  \begin{center}
       \includegraphics[width=0.5\columnwidth]{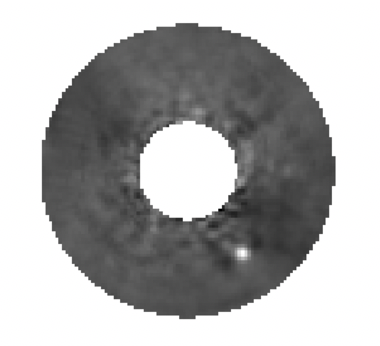}
         \hspace{0.2in}
       \includegraphics[width=1.5\columnwidth]{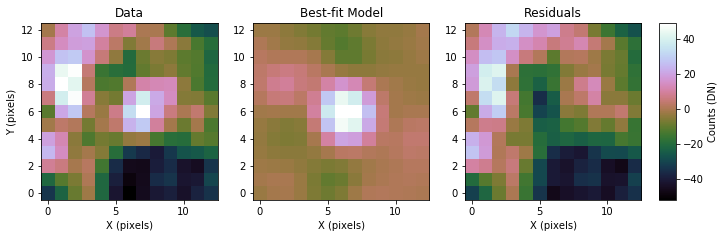}   
   \end{center}
  \caption{(Left) Annulus of KLIP-subtracted $\beta$ Pictoris b GPI data with a synthetic planet  injected at  $pa = 270 \deg$  (3 o'clock) position with $15\%$ of the Forward Model flux. (Right) three-panel plot: (left) search region around the injected planet, (center) the best fit model for $H_{1}$  hypothesis and (right) the residuals after subtracting the best fit model}
  \label{fig:Pictorisb-mock_3_marg_bfm_res}
\end{figure*}

\begin{figure*}
  \begin{center}
       \includegraphics[width=2.0\columnwidth]{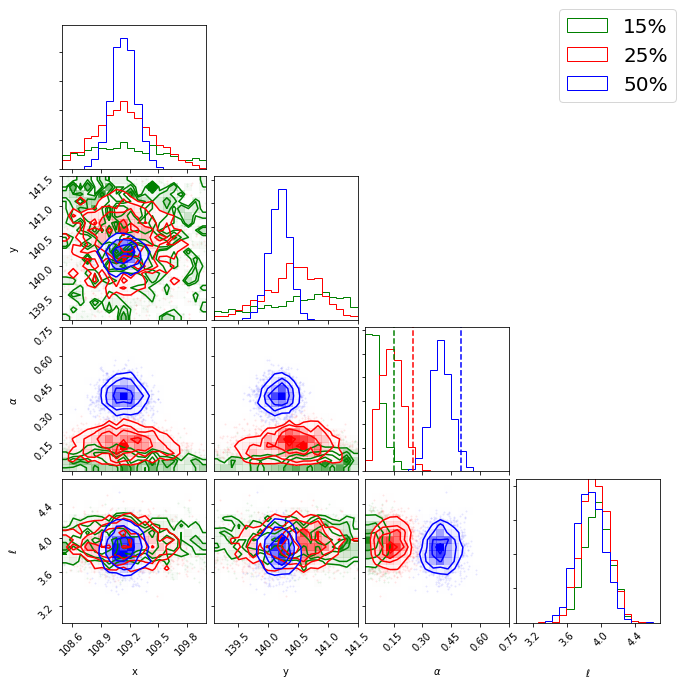} 
   \end{center}
  \caption{The posterior distributions of the four parameters of model $H_{1}$, $(x,y,\alpha,\ell)$, for a planet injected at  $pa = 90 \deg$  (9 o'clock) position with $50\%$ (blue), $25\%$ (red) and $15\%$ (green) of the Forward Model flux.}
  \label{fig:Pictorisb-mock_9_marg_H1}
\end{figure*}

\begin{figure*}
  \begin{center}
       \includegraphics[width=2.0\columnwidth]{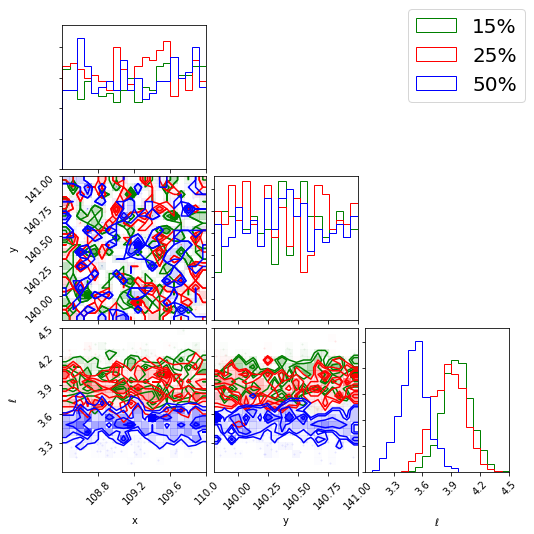}  
   \end{center}
  \caption{The posterior distributions of the three parameters of model $H_{0}$ (null hypothesis), $(x,y,\ell)$, for a planet injected at  $pa = 90 \deg$ (9 o'clock) position with $50\%$ (blue), $25\%$ (red) and $15\%$ (green) of the Forward Model flux.}
  \label{fig:Pictorisb-mock_9_marg_H0}
\end{figure*}

\begin{figure*}
  \begin{center}
        \includegraphics[width=0.5\columnwidth]{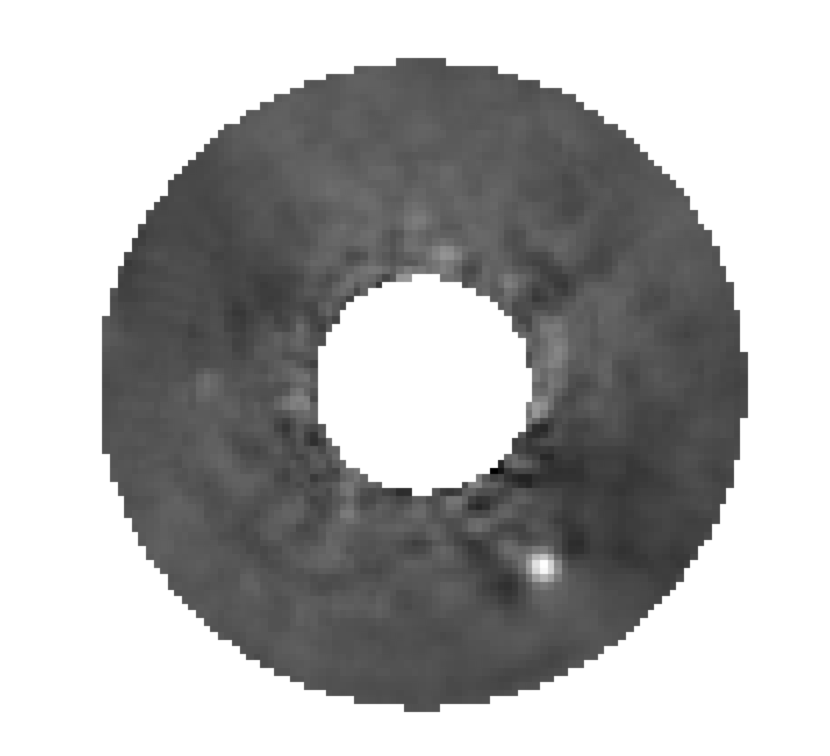}
        \hspace{0.2in}
        \includegraphics[width=1.5\columnwidth]{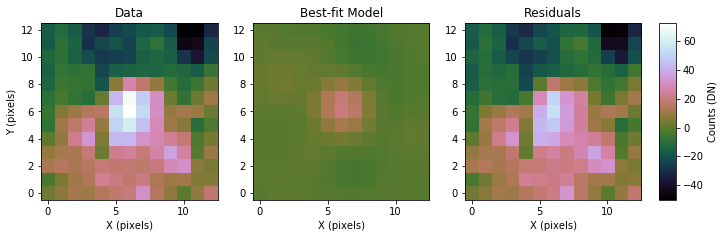}
   \end{center}
  \caption{(Left) Annulus of KLIP-subtracted $\beta$ Pictoris b GPI data with a synthetic planet injected at  $pa = 90 \deg$ position with $15\%$ of the Forward Model flux. (Right) three-panel plot: (left) search region around the injected planet, (center) the best fit model for $H_{1}$ hypothesis and (right) the residuals after subtracting the best fit model.}
  \label{fig:Pictorisb-mock_9_marg_bfm_res}
\end{figure*}

\begin{figure*}
  \begin{center}
  \hbox{
            \includegraphics[width=\columnwidth]{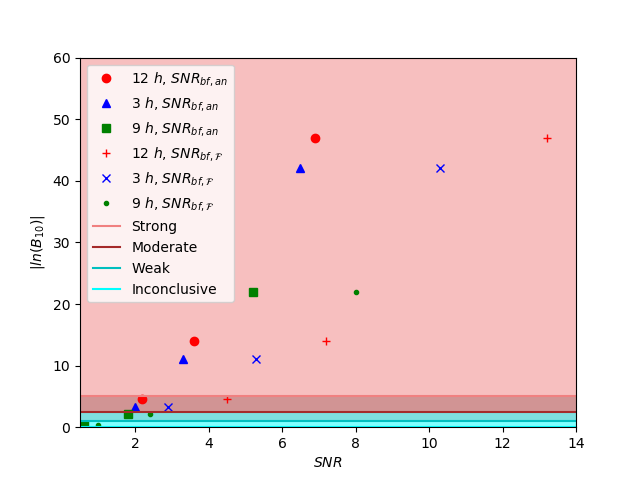}
             \includegraphics[width=\columnwidth]{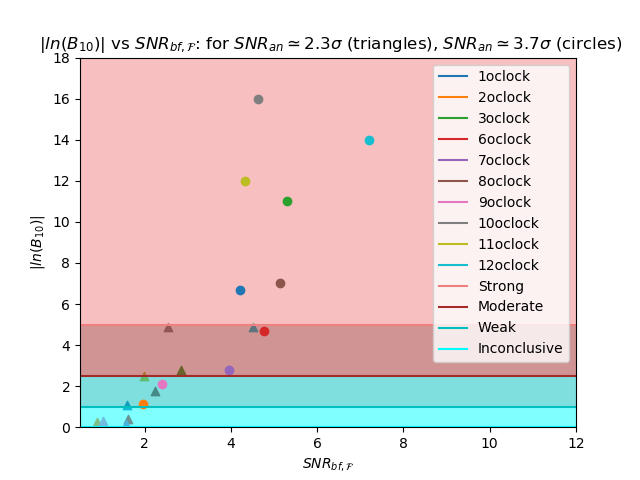}}
   \end{center}
  \caption{Evidence vs SNR: SNR estimated in the annulus, $SNR_{an}$,  after masking $\beta$ Pictoris b and the injected planet; and SNR estimated in the fitting area, $SNR_{bf,\mathcal{F}}$ with noise estimated after subtracting the best fit model from the fitting area. The horizontal lines represent the threshold values which are empirically set, and they occur for values of the logarithm of the Bayes factor of $|ln B10| = 1.0$, 2.5 and 5.0. Shaded areas represent the different levels of evidence above these thresholds according to convention in \citet{trotta08}. }
 \label{fig:Evidence-SNR}
\end{figure*}

For the sake of illustration we start by presenting detailed results for three locations in the image, with parallactic angle $pa = 0 \deg$ (approximately located at 12 o'clock position on the image), $pa = 270 \deg$ ($\simeq$ 3 o'clock) and $pa = 90 \deg$ ($\simeq$ 9 o'clock) followed by results for a larger sample of sources.
All flux injection values are with respect to an initial flux of $5 \times 10^{-5}$ 
 (this is an arbitrary value, approximately that of $\beta$ Pictoris b, prior to converting to contrast units for each wavelength slice).
We inject a planet in each of these locations with decreasing SNR by reducing the flux of the source down to $50\%$ , $25\%$ ,  $15\%$ of the input flux.
To estimate the SNR, the noise is calculated after masking out both the synthetic planet and $\beta$ Pictoris b (we mask a radius of 5 pixels centered at $\beta$ pic b and another 5 pixels centered in the injected planet).
The marginal distributions of the model parameters for these three positions at the progressively dimmer fluxes, the best fit models, and the residuals are plotted in Figures~\ref{fig:Pictorisb-mock_12_marg_H1} -- \ref{fig:Pictorisb-mock_9_marg_bfm_res}.
%
While the logarithm of the evidence ratios as function of the SNR of the injected planets along with the strength of the evidence in favor of model $H_{1}$ are plotted on the left hand side of Figure~\ref{fig:Evidence-SNR}.
Table~\ref{mock-table} gives a summary of these results.

First, we consider $pa = 0 \deg$ position.
Results are shown in Figures~\ref{fig:Pictorisb-mock_12_marg_H1} , \ref{fig:Pictorisb-mock_12_marg_H0}, and \ref{fig:Pictorisb-mock_12_marg_bfm_res}
and tabulated in Table~\ref{mock-table}.
We start by injecting a planet at the $50\%$ level, corresponding to an $SNR_{an}$ = 7.5$\sigma$ and an $SNR_{bf,\mathcal{F}}$ = 13.2$\sigma$.
 The posterior distributions of the source parameters plotted in Figure~\ref{fig:Pictorisb-mock_12_marg_H1} show that the run for $H_{1}$ model converge to the true location of the injected planet in the image. In contrast the posteriors distributions for the position parameters from the $H_{0}$ run are flat and have wide uncertainties (on the order of about $\pm 3$ pixels, see Figure~\ref{fig:Pictorisb-mock_12_marg_H0}).
The logarithm of the evidence ratios, is $\approx 47$  strongly preferring $H_{1}$ over the null hypothesis. 
Next, in the same location, we inject a planet, this time at the $25\%$ level, giving an $SNR_{an}$ = 4.3$\sigma$ and an $SNR_{bf,\mathcal{F}}$ = 7.2$\sigma$. In this case the logarithm of the evidence ratios is 14, indicating still a strong evidence in favor of $H_{1}$ hypothesis. Finally, lowering the injected flux to $15\%$, we recover an evidence ratio of $\approx 4.6$, corresponding to moderate-to-strong evidence in favor of $H_{1}$. 

\begin{table*}
\caption{The PSF fractions, SNR, the logarithm of the evidence ratios and the strength of the evidence in favor of model $H_{1}$ (planet model).}
\begin{center}
\begin{tabular}{|c|c|c|c|c|c|c|c|}
\hline
Position = pa ($\deg$)	& $PSF_{f}$ 	& $SNR_{bf,\mathcal{F}} (\sigma))$ 	& $SNR_{bf,an} (\sigma)$ 	& $SNR_{in,an} (\sigma)$ 	& $SNR_{an} (\sigma)$ 	&$ln(B_{10})$ & Strength \\
\hline
$  0$ 				 &0.50 		&13.2 			&6.9			&7.5			&7.5		&47 			& Strong \\
$ 90 $   				&0.50 		& 8.0		       		&5.2			&7.4			&6.8		&22 			& Strong \\
$ 270 $ 			 	&0.50		&10.3	        		&6.5			&7.3			&5.5		&42 			& Strong \\
$ 0 $      				& 0.25 		&7.2				&3.6			&3.8			&4.3		&14 			& Strong \\
$ 90 $    				&0.25 		&2.4				&1.8		        &3.7		        &3.5		&2.1 			& Weak \\
$ 270 $  				&0.25       		&5.3				&3.3			&3.7			&2.6		&11			& Strong \\
$  0 $				&0.15 		&4.5				&2.2			&2.3			&3.0		&4.6 			& Moderate \\
$ 90 $ 				&0.15 		&1.0				&0.6			&2.2			&2.2		&-0.3 		& Inconclusive \\
$  270 $   				&0.15		&2.9				&2.0			&2.2			&1.5		&3.3  		& Moderate\\
\hline
\end{tabular}
\end{center}
\label{mock-table}
\end{table*}

Next we inject a synthetic planet in a different location, now at $pa = 270  \deg$ position.
Results are shown in Figure~\ref{fig:Pictorisb-mock_3_marg_H1}, \ref{fig:Pictorisb-mock_3_marg_H0} and \ref{fig:Pictorisb-mock_3_marg_bfm_res},
and tabulated in Table~\ref{mock-table}.
 Once again we start by injecting a planet at the $50\%$ level, corresponding to $SNR_{an}$ = 5.5$\sigma$ and an $SNR_{bf,\mathcal{F}}$ = 10.3$\sigma$. The logarithm of the evidence ratios is $42$, indicating a strong evidence for hypothesis $H_{1}$.
 When we reduce the injected flux to $25\%$, $SNR_{an}$ = 2.6$\sigma$ and an $SNR_{bf,\mathcal{F}}$ = 5.3$\sigma$, and the logarithm of the evidence ratios is 11, indicating still a strong evidence for $H_{1}$ model. 
 Next, we reduce the injected planet flux to $15\%$, with an $SNR_{an}$ = 1.5$\sigma$ and an $SNR_{bf,\mathcal{F}}$ = 2.9$\sigma$. The logarithm of the evidence ratios is 3.3 indicating moderate evidence at this location.

 Finally we consider the position at  $pa = 90 \deg$.
 Results are shown in Figure~\ref{fig:Pictorisb-mock_9_marg_H1},  \ref{fig:Pictorisb-mock_9_marg_H0} and \ref{fig:Pictorisb-mock_9_marg_bfm_res} 
 and tabulated in Table~\ref{mock-table}.
 Once again we start by injecting a planet with $SNR_{an}$ = 6.8$\sigma$ and an $SNR_{bf,\mathcal{F}}$ = 8.0$\sigma$. The logarithm of the evidence ratios is $22$, indicating a strong evidence for hypothesis $H_{1}$.
 When we reduce the injected flux to $25\%$, $SNR_{an} = 3.5 \sigma$ and an $SNR_{bf,\mathcal{F}} = 2.4 \sigma$. the logarithm of the evidence ratios is 2.1, indicating moderate evidence for the $H_{1}$ model. 
 Finally we inject a fainter planet in the same location with an $SNR_{an} = 2.2 \sigma$ and an $SNR_{bf,\mathcal{F}} = 1.0 \sigma$. In this location, the $15\%$ flux corresponds to an evidence ratio of $0.3$, indicating no evidence in favor of $H_{1}$. Thus, in this location at this dim flux level, the resulting low SNR means the true planet cannot be distinguished from the surrounding background noise.

 This simple exercise indicates that a planet with flux above a threshold of the order of $SNR_{an} \simeq 2 \sigma$  can, in principle,  be detected relatively confidently using evidence ratios.
 From Figure~\ref{fig:Evidence-SNR} the resulting SNR and evidence follow a similar trend for all cases. However the $pa = 90 \deg$ (9 o'clock) position shows systematically lower SNR for same fraction of the input flux injected and lower evidence ratios. 
The marginal distributions of the parameter $\alpha$  and  the residual plots for the fainter injected planets plotted in Figures~\ref{fig:Pictorisb-mock_12_marg_H1}, \ref{fig:Pictorisb-mock_12_marg_bfm_res}, \ref{fig:Pictorisb-mock_3_marg_H1},  \ref{fig:Pictorisb-mock_3_marg_bfm_res}, \ref{fig:Pictorisb-mock_9_marg_H1}, and \ref{fig:Pictorisb-mock_9_marg_bfm_res}  show that the  planet's flux is slightly overestimated for the $pa = 0 \deg$ (12 o'clock) position, underestimated for the  $pa = 90 \deg$  (9 o'clock) and reasonably recovered ie unbiased for the $pa = 270 \deg$ (3 o'clock) location. 

Planet's flux overestimation is common if the injected planet falls on the top of a speckle. 
Underestimation can occur when part of the planet's flux is subtracted when removing the residual speckles from the image in the KLIP step. 
This might happen if the chosen number of KL modes is insufficient to account for the noise characteristics and/or the KLIP procedure is insufficient (eg. the reference frames chosen are inadequate) to properly separate both contributions, or if there is insufficient sky rotation.
Mistakenly some of the planet's flux ends up contributing to the KLIP modes describing the residual speckles. The planet's flux is thus partially subtracted when the KLIP modes describing the speckles are subtracted from the image resulting in a underestimation of the true planet's flux rendered by the parameter $\alpha$. 
KLIP-FM should be compensating for these effects by accounting for the distortions of the original planet PSF due to the KLIP step.
Hence this mild underestimation might indicate that the forward-model does not completely account for the planet's PSF distortions at this position on the image. This or the log-uniform prior on  the flux is dictating the fits as the $SNR_{bf,an}$ and $SNR_{bf, \mathcal{F}}$ are systematically low (and lower then the SNR on the other positions for the same planet's input flux). 

We also note that the local $SNR_{bf,\mathcal{F}}$ is systematically higher then the $SNR_{bf,an}$ for the three locations.
 The evidence values and SNR differences at each location for each injected flux, shows that the evidence for $H_{1}$ depends on the local noise estimated in the fitting region.
 It appears that the local fitted noise is systematically lower then the overall noise estimated in the annulus. This could be due to the presence of a higher speckle residual noise in the inner boundary of the annulus as this region is not covered by the fitting regions around the selected locations of the injected planets.

To explore this further we inject a few sources at different locations in the annulus (at 1, 2, 3, 4, 6, 7, 8, 9, 11, 12 o'clock positions) considering 0.15 and 0.25 percentages of the FM which correspond to a SNR estimated in the annulus $SNR_{an} \simeq 2\sigma$ and $SNR_{an} \simeq 4\sigma$ respectively.
The parallactic angle is measured in 30 degree intervals counterclockwise, so 0 degrees is 12 o'clock, 30 degrees is 11 o'clock, so on, 300 degrees is 2 o'clock, 330 degrees is 1 o'clock. 

On the right hand side of Figure~\ref{fig:Evidence-SNR} we plot the $|ln(B10)|$ as function of the SNR estimated in the local fitting area $SNR_{bf,\mathcal{F}}$ for the several injected sources.
We start by noting that the $SNR_{bf,\mathcal{F}}$ deviates from $SNR_{bf,an}$ and the $SNR_{an}$. The differences with $SNR_{an}$ can either indicate that the background noise is more complex then that described by a Matern covariance matrix (the correlation length $\ell$ varies from 3 to 6, the expected value is 3) and/or there is a over or under estimation of the planet flux. 
For example, the $25$ and $15$ percentages of the planet's FM injected at $pa = 90 \deg$ (9 o'clock) position exhibit lower $SNR_{bf,\mathcal{F}}$ then $SNR_{an}$. This is in agreement with the observed underestimation of its flux shown in Figure~\ref{fig:Pictorisb-mock_9_marg_H1}. 
While the flux of the planet injected at $ pa =  270 \deg$ (3 o'clock) position is reasonably recovered indicating that differences between these two SNR estimators might be due to deficiencies in the noise characterization.
Indeed there are larger scale correlations of noise in the image that we are ignoring because we are fitting a small local fitting region where that large-scale noise is $\simeq$ constant. 

\begin{figure*}
  \begin{center}
\hbox{
      \includegraphics[width=\columnwidth]{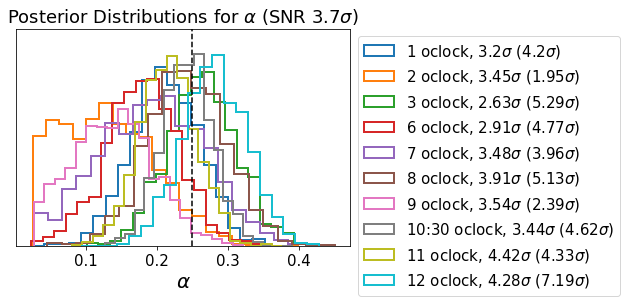}
     \includegraphics[width=\columnwidth]{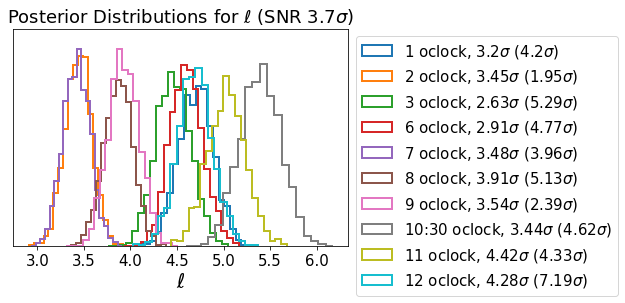}}    
   \end{center}
  \caption{Left: Posterior distributions of $\alpha$; Right: Posterior distributions of $\ell$ for all injected sources (ie for the several positions on the image) for $SNR_{in,an} \simeq 3.7 \sigma$; $f_{FM_{psf}} =0.25$}
  \label{fig:marginals-alpha-ell-025}
\end{figure*}

\begin{figure*}
  \begin{center}
\hbox{
     \includegraphics[width=\columnwidth]{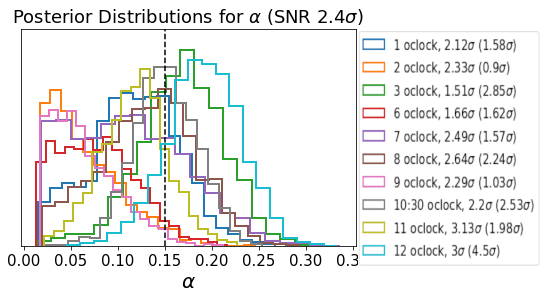}
     \includegraphics[width=\columnwidth]{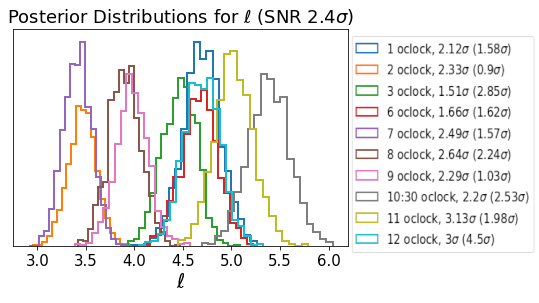}}
   \end{center}
  \caption{Left: Posterior distributions of $\alpha$; Right: Posterior distributions of $\ell$ for all injected sources (ie for the several positions on the image) for $SNR_{in,an} \simeq 2.4 \sigma$; $f_{FM_{psf}} =0.15$}
  \label{fig:marginals-alpha-ell-015}
\end{figure*}

Figure~\ref{fig:marginals-alpha-ell-025} and Figure~\ref{fig:marginals-alpha-ell-015} show the posterior distributions of the source parameter $\alpha$ and the correlated residual noise (speckle) parameter $\ell$ for all positions in the image for $SNR_{an} \simeq 3.7 \sigma$; $f_{FM_{psf}} =0.25$ and $SNR_{an} \simeq 2 \sigma$; $f_{FM_{psf}} =0.15$ respectively. 
The marginal distributions of the parameter $\alpha$ indicate a clear flux underestimation by more than $1\sigma$ for four of the ten locations in the image (clearly so in Figure~\ref{fig:marginals-alpha-ell-015}), with skewed distributions towards lower values of $\alpha$. 
The posterior distributions are skewed to lower values when the SNR is low. This seems to indicate that the prior distribution is influencing the posterior distribution (as mentioned above). Hence this underestimation could, in principle, also be due to the choice of a log-uniform prior on the flux, rather than a uniform prior.

There is no clear correlation with the best fit  $\ell$ values.
The peaks of the marginal distributions of the parameter $\ell$ fall in the range  $3\leq \ell \leq 6$. 
One would expect the distribution to peak around $\ell \sim 3$.
This indicates the noise correlation length can be larger then expected. We are probably seeing larger scale correlations in some of the fitting regions such as the AO wind-butterfly pattern.

In this paper we consider Gaussian Likelihoods and therefore we are not accounting for the potential non-Gaussianity of the real noise. This miss-modelling of the statistics of the background noise can impact the estimation of the evidence.
For example, in some cases, the fat tails of the non-Gaussian noise might give rise to a stronger evidence then it is in reality. 
The severity of this depends on how non-Gaussian the background noise is.  In future work we will assess the resulting bias due to assuming a Gaussian noise. To this end we will construct non-Gaussian Likelihoods following approaches described in \citet{rocha-ng-01,rocha-ng-05}.


Finally we tested this approach by running {\tt PlanetEvidence} on "noise", that is, when no synthetic planet is injected into the data.

\begin{figure*}
  \begin{center}          
  	\includegraphics[width=2.0\columnwidth]{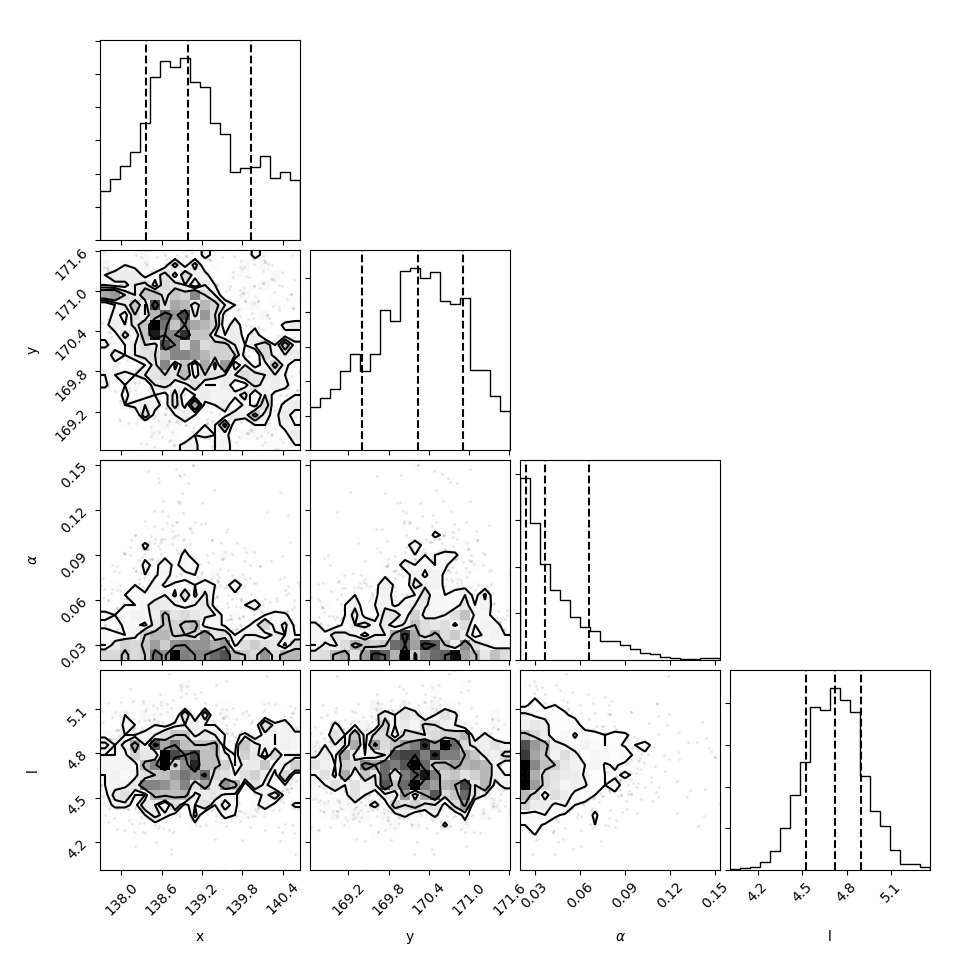}
    \end{center}         
  \caption{The posterior distributions of the four parameters of model $H_{1}$, $(x,y,\alpha,\ell)$ for the runs of {\tt PlanetEvidence} on noise ie when no synthetic planet is injected into the data, at  $pa= 0 \deg$ (12 o'clock) position in the image. }
    \label{fig:Pictorisb-noise_12_H1}
\end{figure*}

\begin{figure*}
  \begin{center}
      	 \includegraphics[width=2.0\columnwidth]{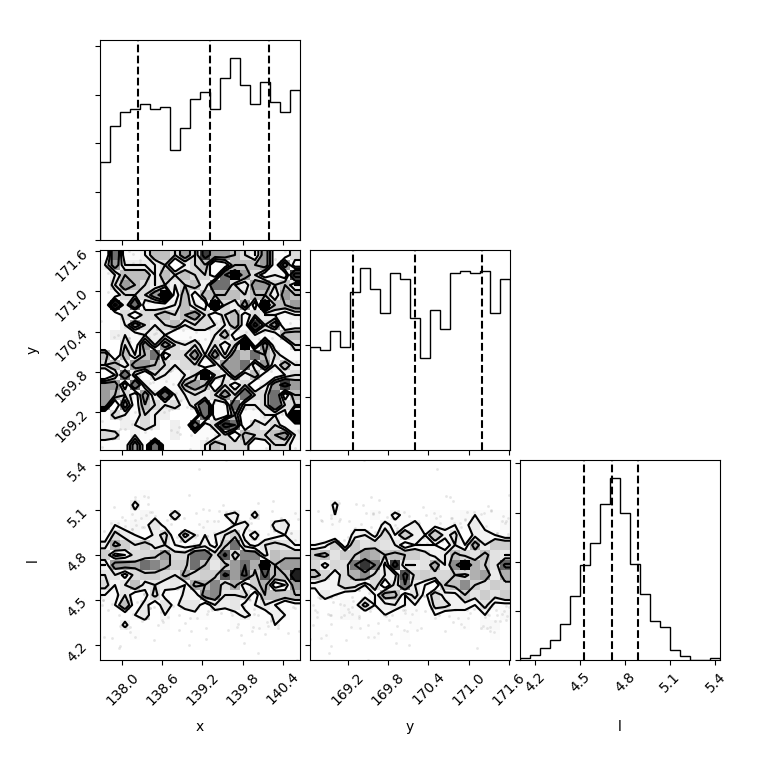}   
   \end{center}
  \caption{The posterior distributions of the three parameters of model $H_{0}$ (null hypothesis), $(x,y,\ell)$ for the runs of {\tt PlanetEvidence} on noise ie when no synthetic planet is injected into the data, at $pa= 0 \deg$ (12 o'clock) position in the image. }
    \label{fig:Pictorisb-noise_12_H0}
\end{figure*}

\begin{figure*}
  \begin{center}
          \includegraphics[width=2.0\columnwidth]{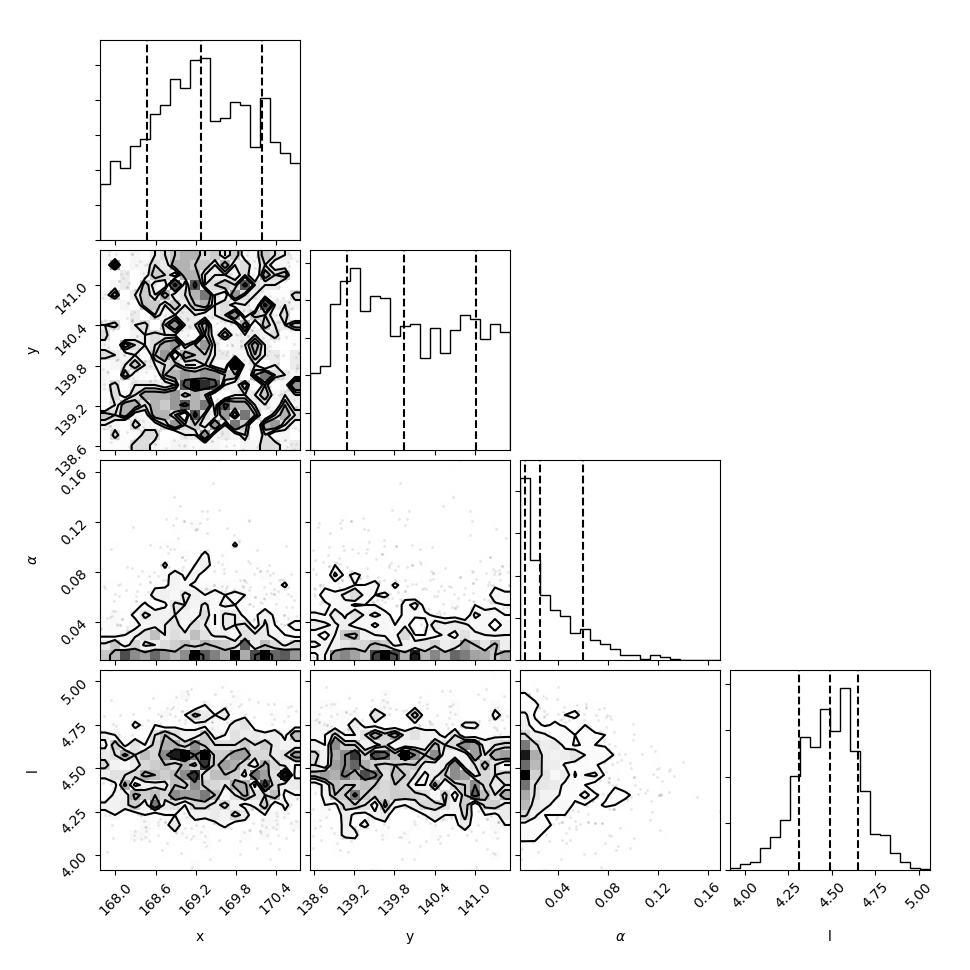}
   \end{center}          
  \caption{The posterior distributions of the four parameters of model $H_{1}$, $(x,y,\alpha,\ell)$ for the runs of {\tt PlanetEvidence} on noise ie when no synthetic planet is injected into the data, at  $pa= 270 \deg$ (3 o'clock) position in the image. }
    \label{fig:Pictorisb-noise_3_H1}
\end{figure*}

\begin{figure*}
  \begin{center}
       \includegraphics[width=2.0\columnwidth]{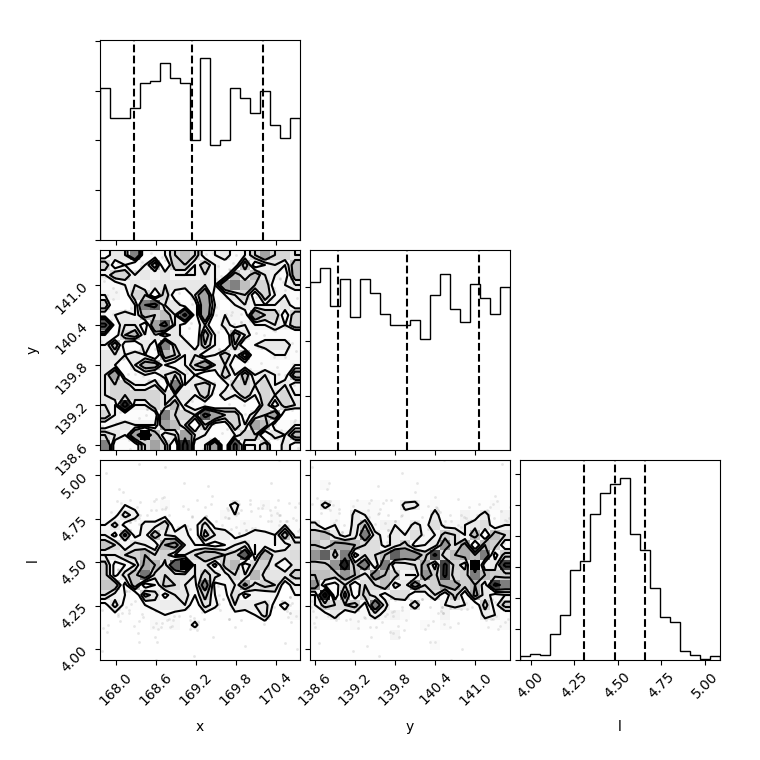}   
   \end{center}
  \caption{The posterior distributions of the three parameters of model $H_{0}$ (null hypothesis), $(x,y,\ell)$ for the runs of {\tt PlanetEvidence} on noise ie when no synthetic planet is injected into the data, at $pa= 270 \deg$ (3 o'clock) position in the image. }
    \label{fig:Pictorisb-noise_3_H0}
\end{figure*}

\begin{figure*}
  \begin{center}
          \includegraphics[width=2.0\columnwidth]{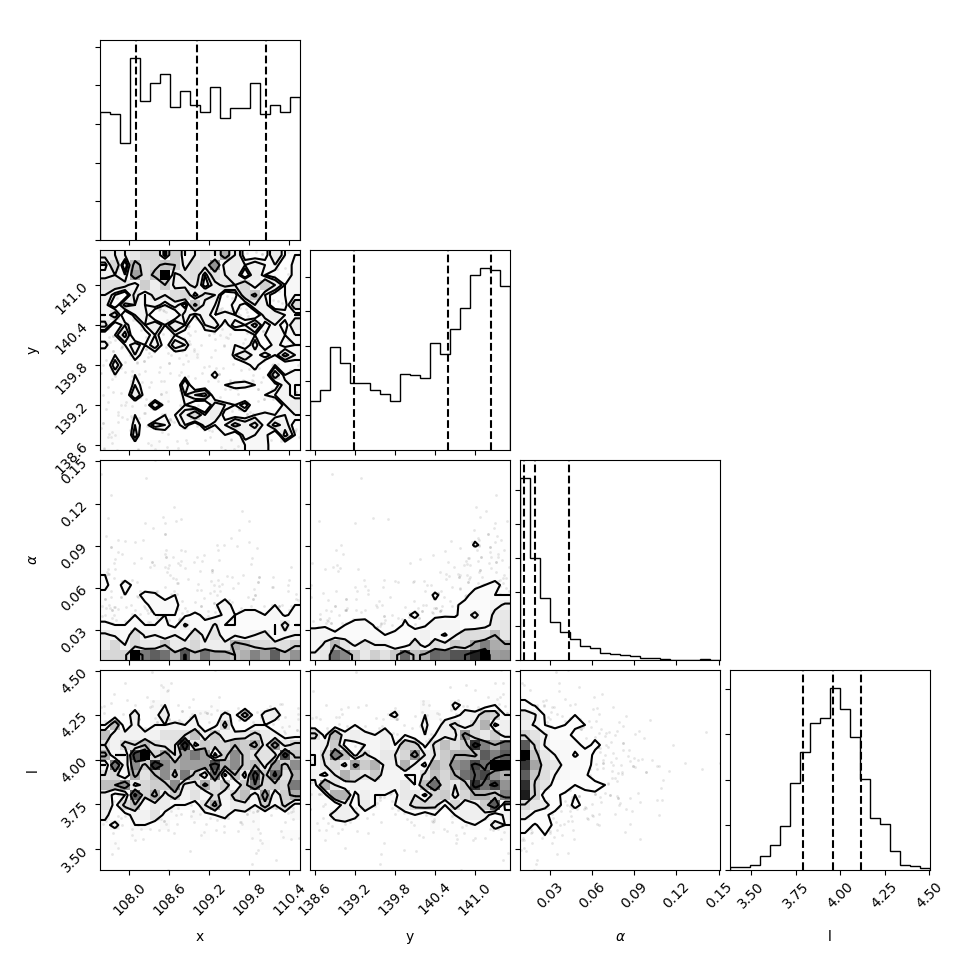}
   \end{center}       
  \caption{The posterior distributions of the four parameters of model $H_{1}$, $(x,y,\alpha,\ell)$ for the runs of {\tt PlanetEvidence} on noise ie when no synthetic planet is injected into the data, at  $pa= 90 \deg$ (9 o'clock) position in the image. }
    \label{fig:Pictorisb-noise_9_H1}
\end{figure*}

\begin{figure*}
  \begin{center}
       \includegraphics[width=2.0\columnwidth]{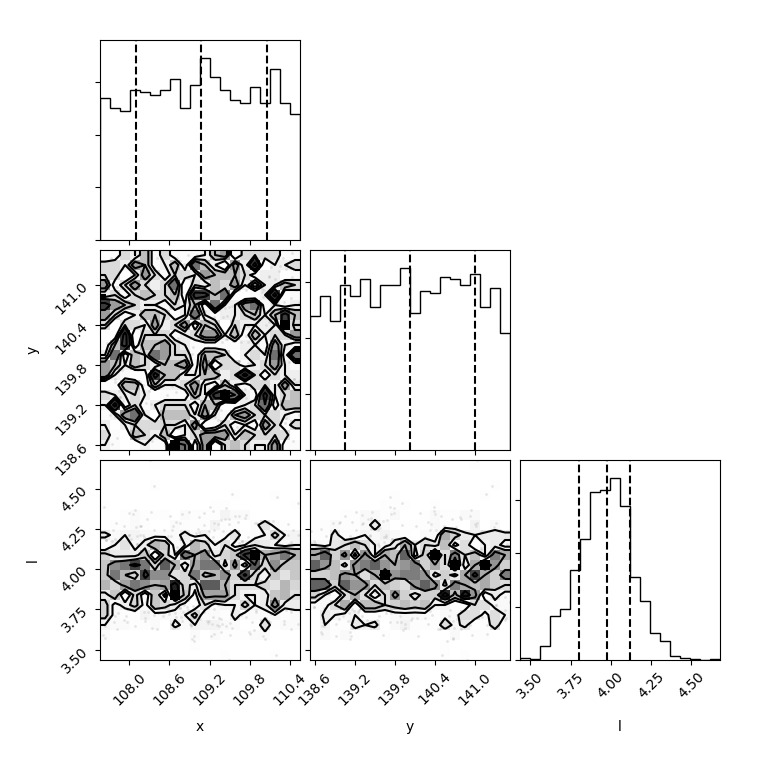}   
   \end{center}
  \caption{The posterior distributions of the three parameters of model $H_{0}$ (null hypothesis), $(x,y,\ell)$ for the runs of {\tt PlanetEvidence} on noise ie when no synthetic planet is injected into the data, at $pa= 90 \deg$ (9 o'clock) position in the image. }
    \label{fig:Pictorisb-noise_9_H0}
\end{figure*}

\begin{figure*}
  \begin{center}
  \vbox{
        \includegraphics[width=2.0\columnwidth]{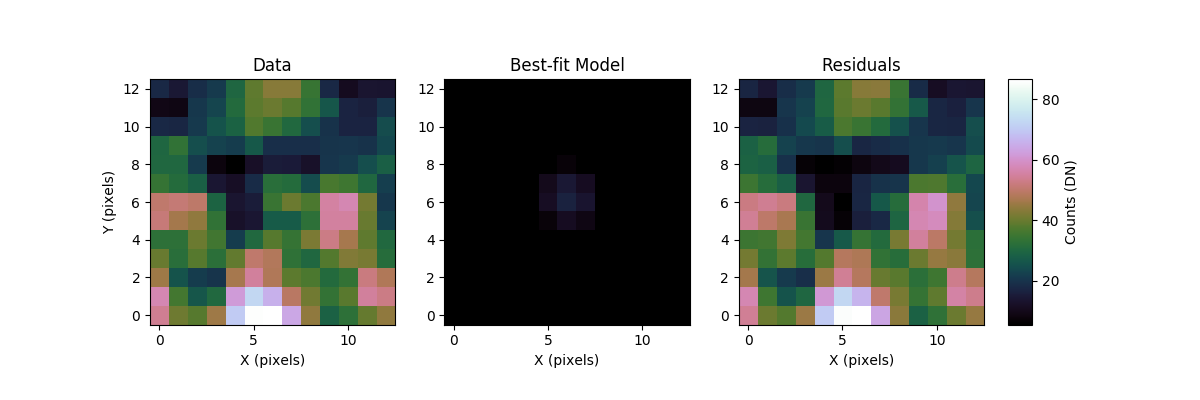}
        \includegraphics[width=2.0\columnwidth]{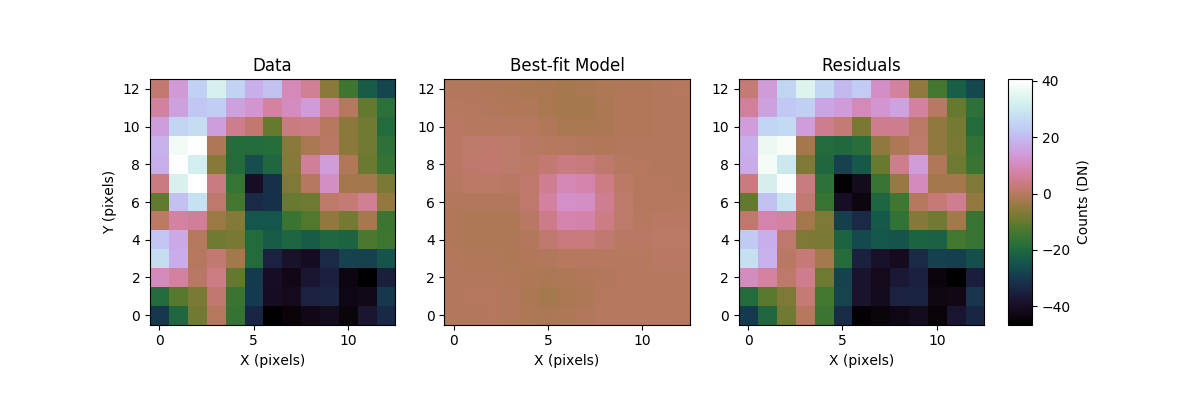}
          \includegraphics[width=2.0\columnwidth]{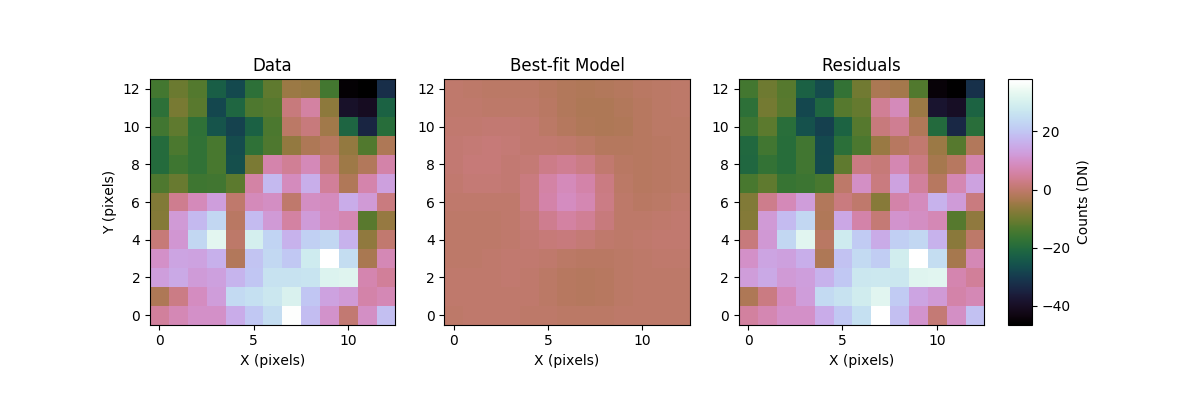}}     
   \end{center}
  \caption{Three panel plot: (left) search region around the chosen position on the image, (center) the best fit model for $H_{1}$ hypothesis and (right) the residuals after subtracting the best fit model; (Top) for $pa=0 \deg$ (12 o'clock); (Middle) for  $pa=270 \deg$ (3 o'clock) and (bottom) for $pa=90 \deg$ (9 o'clock)}
  \label{fig:Pictorisb-noise_marg_bfm_res}
\end{figure*}

We consider the three $pa = 0 \deg$, $pa=90 \deg$ and $pa=270 \deg$  positions in the image. Results are shown in Figures~\ref{fig:Pictorisb-noise_12_H1} -- \ref{fig:Pictorisb-noise_marg_bfm_res}, while the results are tabulated on Table~\ref{noise-table} for all positions in the image.

We find that seven out of the ten cases considered hold $B10 \approx 0.3$, using Harold Jeffreys scale interpretation for $B_{10}$ this indicates that Evidence supports the null hypothesis, $H_{0}$ \citet{jeffreys61}, while the remaining three cases are inconclusive.

\begin{table*}
\caption{The SNR, the logarithm of the evidence ratios and the strength of the evidence in favor of model $H_{0}$ (null hypothesis) when no synthetic planet is injected into the data.
A negative (positive) value for $ln(B_{10})$  indicates that the competing model is disfavoured (supported) with respect to the null hypothesis $H_{0}$. Seven out of ten cases support the null hypothesis, $H_{0}$.}
\begin{center}
\begin{tabular}{|c|c|c|c|c|c|}
\hline
Position=pa ($\deg$)	  & $SNR_{bf,\mathcal{F}} (\sigma))$ 	& $SNR_{an} (\sigma)$ 	&$ln(B_{10})$	 &B10  & Strength \\
\hline
$  0 $ 		&1.1							&1.7						&-1.2				& 0.3			& Weak\\
$90 $ 		& 0.5		       					&1.1						&-1.3				& 0.3		 	& Weak \\
$270 $		&0.6	        						&0.4						&-0.9				& 0.4			& Inconclusive \\
$30 $		&0.574						&2.061		   			&-1.4	        			&0.3			& Weak \\
$60 $ 		&0.310						&-0.702					&0.004        			&1.0			& Inconclusive \\
$120 $	        &0.690						&1.199					&-0.7	      			&0.5                 	& Inconclusive \\
$150 $		&0.683						&1.758					&-1.2	      			&0.3               	& Weak\\
$180 $		&0.594						&0.945					&-1.1	      			&0.3                	& Weak \\
$300  $	         &0.488                                               &1.395                                       &-1.3               			&0.3                	& Weak \\
$330 $	         &0.553                                               &1.045             			        &-1.3                 			&0.3               	& Weak\\ 
\hline
\end{tabular}
\end{center}
\label{noise-table}
\end{table*}
%

In order to prove that the false positive rate is low enough to confidently say they are planets a more thorough study is required.
In a future publication we will present such study by testing several different positions on noise only simulations (not that here we use the observed data itself).
As it is our results indicate that using Evidence ratios we can detect 'real' sources otherwise not seen by eye in the image. 

Although KLIP-FM allows for accurate astrometry of a potential object, it does so given an initial guess of the point source location (for example if detected by eye).
Here we have shown that our method can relax this precondition, obviating the implementation of a blind detection step. 
This step does not use matched filters (as in Forward Model Matched Filter (FMMF) module in {\tt pyKLIP} described in \citet{ruffio17}), but rather incorporate all the information in the data model and Likelihood and marginalize over the nuissance parameters. 
When implemented in the coadded data this step entails searching for planets in all pixels in the image and constructing a catalog of planets ordered according to the respective evidence value (following a similar procedure presented in \citet{rocha09,rocha12}.

\section{Conclusions}

In this paper we introduced a methodology to determine whether a planet detected using KLIP-FM is a true point source rather than a residual background noise. 
This is achieved by constructing two models: $H_{1}$ - a "planet is present" and $H_{0}$ -  null hypothesis ("no planet" present), and using nested sampling to compute the evidence for each model. We test this methodology by forward modeling on the location of $\beta$ Pictoris b and computing the corresponding evidence ratio for the $H_{1}$ and $H_{0}$ models. As expected, we get extremely strong evidence in favor for the planet being present rather than background noise. Next we test our approach on synthetic planets injected into the image. The evidence for the  'planet present' hypothesis weakens as the SNR of the injected source decreases. 
We have shown that dim sources, not necessarily seen by eye, can be detected and characterized as true point sources rather than background noise. 
To assess performance against false positives we tested our approach when no planet is injected. As expected, in most of the cases, we get evidence in favor of the null hypothesis.
The exercise  presented here act as a proof of concept for a Bayesian-based algorithm for true blind detection, that is, not requiring an initial guess as to the location of the planet. 
As such, by initializing forward modeling at locations of interest (such as brighter points) across the KLIP-subtracted image, KLIP-FM may converge on a potential true source and the evidence that the detected source is more likely a true point source estimated with the {\tt PlanetEvidence} module.

{\tt PlanetEvidence} is implemented in {\tt pyKLIP} and is run in conjunction with KLIP-FM \footnote {Visit https://pyklip.readthedocs.io/en/latest/ for {\tt PlanetEvidence} release notes, examples and tutorials.}

\newpage
\section{Acknowledgements}
GR would like to acknowledge useful discussions with Jeff Jewell.
The research presented here was carried out at the Jet Propulsion Laboratory, California Institute of Technology, under a contract with the National Aeronautics and Space Administration.

\end{document}